\documentclass[final,letter,5p,times,twocolumn]{elsarticle}

\usepackage{lineno,hyperref}
\biboptions{sort&compress}
\journal{Elsevier}

\usepackage[T1]{fontenc}
\usepackage[latin9]{inputenc}
\usepackage{amsmath}
\usepackage{graphicx}
\usepackage{color}
\usepackage{graphicx}
\usepackage{multirow}
\usepackage{amsmath,bm}

\newcommand {\snn}	{\sqrt{s_{_{\rm NN}}}}

\newcommand {\Nch}	{N_{\rm ch}}

\newcommand {\Zr}	{$^{96}$Zr}
\newcommand {\Ru}	{$^{96}$Ru}
\newcommand {\RuRu}	{$^{96}_{44}$Ru+$^{96}_{44}$Ru}
\newcommand {\ZrZr}	{$^{96}_{40}$Zr+$^{96}_{40}$Zr}
\newcommand {\RuZr}	{Ru+Ru/Zr+Zr}

\newcommand {\rnp}	{$\Delta r_{\rm np}$}
\newcommand {\Rp}	{R_{\rm p}}
\newcommand {\Rn}	{R_{\rm n}}
\newcommand {\Ap}	{a_{\rm p}}
\newcommand {\an}	{a_{\rm n}}
\newcommand {\ecc}	{\epsilon_2}

\newcommand {\bRu}	{\beta_{2}^{\rm Ru}}
\newcommand {\bZr}	{\beta_{2}^{\rm Zr}}

\newcommand {\trento}   {{\sc trento}}

\newcommand {\mean}[1]	{\langle #1\rangle}

\begin{document}

\begin{frontmatter}

\title{Determine the neutron skin type by relativistic isobaric collisions}

	\author[1]{Hao-jie Xu\corref{mycorrespondingauthor}}\ead{haojiexu@zjhu.edu.cn}
\author[2]{Hanlin Li}
\author[1]{Xiaobao Wang}
\author[1]{Caiwan Shen}
	\author[1,3]{Fuqiang Wang\corref{mycorrespondingauthor}}\ead{fqwang@purdue.edu}

\address[1]{School of Science, Huzhou University, Huzhou, Zhejiang 313000, China}
\address[2]{College of Science, Wuhan University of Science and Technology, Wuhan, Hubei 430065, China}
\address[3]{Department of Physics and Astronomy, Purdue University, West Lafayette, Indiana 47907, USA}

\cortext[mycorrespondingauthor]{Corresponding author}

\date{\today}

\begin{abstract}
The effects of neutron skin  on the multiplicity ($\Nch$) and eccentricity($\epsilon_2$)  in relativistic \RuRu\ and \ZrZr\  collisions at $\snn=200$ GeV
	are investigated with the Trento model. It is found that the \RuZr\ ratios of the $\Nch$ distributions and $\epsilon_{2}$ in mid-central collisions are exquisitely sensitive to the neutron skin type (skin vs.~halo). The state-of-the-art calculations by energy density functional theory (DFT) favor the halo-type neutron skin and can soon be confronted by experimental data. 
    It is demonstrated that the halo-type density can serve as a good surrogate for the DFT density, and  thus can be efficiently employed  to  probe nuclear deformities by using elliptic flow data in central collisions. 
    We provide hereby a proof-of-principle venue to simultaneously determine the neutron skin type, thickness, and nuclear deformity.
\end{abstract}

\end{frontmatter}

\section{Introduction}

Nuclear densities are often parameterized by Woods-Saxon (WS) distributions. Proton and
neutron distributions are usually not distinguished and the former, which is well measured, is substituted for the latter. For most studies in heavy ion collisions,
this is sufficient. However, it is well known that proton and neutron
distributions differ because of 
Coulomb interactions and the symmetry energy. This subtle difference becomes
important when one compares collisions of similar species,
such as isobar \RuRu\ and \ZrZr\ collisions~\cite{Xu:2017zcn,Li:2018oec,Hammelmann:2019vwd,Li:2019kkh}, recently conducted at the Relativistic Heavy Ion Collider (RHIC)~\cite{Adam:2019fbq}. Because Ru has more protons than Zr,
the Ru charge distribution is fatter than the Zr's. Simply substituting the charge density
for the mass density, one predicts a lower multiplicity in central Ru+Ru than  Zr+Zr collisions~\cite{Deng:2016knn}. Energy density functional theory (DFT) calculations indicate a significantly thicker neutron skin in \Zr\ than \Ru\ and that
the overall size of  \Zr\  is bigger than \Ru. Because of this, DFT instead predicts
 a higher multiplicity in central Ru+Ru than  ZrZr collisions~\cite{Xu:2017zcn,Li:2018oec}.
It is further found that, even incorporating the smaller Ru than Zr radius,
the WS skin-type density 
parameterization cannot reproduce the DFT prediction~\cite{Li:2018oec}.
In this paper, we demonstrate that the DFT densities can be significantly
better modeled by the halo-type WS parameterization. The skin- and halo-type
densities predict quite different multiplicity  distributions and elliptic flow, which can be easily discriminated by experimental data. 
If the data, soon to be available, favor the halo-type density
(and therefore verifying the DFT calculation), then our study suggests that the halo-type WS density approximation
can be used as a surrogate for DFT density for most applications in heavy ion collisions, including studies of nuclear deformity. This would provide an efficient way for such studies as DFT calculations of deformed nuclei are challenging.

The rest of the paper is organized as follows. 
Section~\ref{sec:setup} gives a brief description of the \trento\ model used in this work. 
Section~\ref{sec:skin} discusses the effects of nuclear densities, focusing on the difference of neutron skin types and effect of the DFT-calculated densities. 
In Sec.~\ref{sec:deformity}, the effect of nuclear deformity is investigated.
A summary is given in Sec.~\ref{sec:summary}.  

\section{Model Setup}\label{sec:setup}

In relativistic heavy ion collisions, the charged hadron multiplicity ($\Nch$) distribution and elliptic flow ($v_2$) are  two basic observables. For  $v_{2}$, one usually relies on  
macroscopic hydrodynamic or microscopic transport
model calculations; however, those calculations are time-consuming.  
In this study, we
use the \trento\ model~\cite{Bernhard:2016tnd,Moreland:2014oya} to simulate  Ru+Ru collisions and Zr+Zr collisions at nucleon-nucleon center-of-mass energy $\snn=200$ GeV~\cite{Li:2019kkh}.
In \trento\ the transverse area entropy density is proportional to the reduced thickness,
\begin{equation}\label{eq:s}
s(x,y) \propto [(T_{A}^{p} + T_{B}^{p})/2]^{1/p}\,,
\end{equation}
where $T_A=\int \rho(\mathbf{r}) dz$  is the  nuclear thickness function. 
Particle production yield is proportional to the total entropy, 
$\langle\Nch\rangle \propto \int s(x,y)dxdy$, and a  Poisson-type fluctuation is used to generate the event-by-event $\Nch$.
We use the parameter $p=0$ (i.e., $s\propto\sqrt{T_AT_B}$), a gamma fluctuation parameter $k=1.4$,
and a Gaussian nucleon size of $0.6$ fm, which were found to well describe the multiplicity data in heavy ion collisions~\cite{Bernhard:2016tnd,Moreland:2014oya}.
The $v_{2}$ is approximately proportional to the initial eccentricity~\cite{Ollitrault:1992bk},  $\ecc$, at small amplitude ($\ecc<0.5$)~\cite{Noronha-Hostler:2015dbi}.
The conversion (proportionality) factor depends on dynamics and is expected to be the same for the highly similar \RuRu\ and \ZrZr\  collisions at $\snn=200$ GeV. Since in this study we focus only on the relative difference between the two  systems, we will simply compute 
$\ecc^{\rm RuRu}\{2\}/\ecc^{\rm ZrZr}\{2\}$, and substitute that for the $v_2$ ratio.
The $\ecc$ can be readily obtained 
from the initial geometry in the \trento\ model by~\cite{Poskanzer:1998yz}
\begin{align}
	\epsilon_{2} & =\frac{\int r^{2}e^{in\phi}s(x,y)dxdy}{\int r^{2}s(x,y)dxdy}\,,\\
	\epsilon_{2}\{2\} & =\sqrt{\langle\epsilon_{2}^{2}\rangle}\,,
\end{align}
where $\phi$ is the azimuthal angle of position $(x,y)$ and $\langle\cdots\rangle$ denotes the event average. 

We note that, while we focus on the \trento\ model, there are several initial geometry models on market. The most traditional one is the two-component Glauber model where the $\Nch$ depends on the numbers of participants and binary collisions~\cite{Kharzeev:2000ph,Kharzeev:2001yq,Alver:2006wh,Miller:2007ri,Xu:2014ada}. The essential difference of all those models lies in how $\Nch$ or $s$ depends on the thickness function (Eq.~(\ref{eq:s})). 
The treatment of fluctuations in those models can also affect the magnitude of $\ecc$  in each collision system, as well as the relation between $\Nch$ and centrality~\cite{Miller:2007ri,Alvioli:2011sk,Rybczynski:2011wv,Blaizot:2014wba,Alvioli:2018jls}.
Since we are primarily interested in the relative difference between the isobar systems, these model dependencies are  weak.
Transport models are more complex, but the essential ingredient relevant for our study is, again, the effective relationship of Eq.~(\ref{eq:s}).
Studies have previously been carried out by a multi-phase transport (AMPT) model~\cite{Li:2018oec}, and consistent results have been obtained as we will refer to along with our results.

\section{Effects of Nuclear densities}\label{sec:skin}
Besides some model dependence in $\Nch$, the calculation boils down to the thickness function $T_A$. Different nuclear densities yield different $T_A$; thus the $\Nch$ distribution and $v_2$ can be used to determine the nuclear density. Using the relative measures of isobar systems reduces the large uncertainties in theoretical modeling of the collision dynamics. Our starting point is the measured charge distributions of Ru and Zr from experimental data~\cite{Fricke:1995zz}.

\subsection{WS charge densities}
The charge densities of Ru and Zr 
were parameterized by WS distribution (also known as two-parameter Fermi distribution), 
\begin{equation}
	\rho(r)=\frac{\rho_{0}}{1+\exp\left[(r-R)/a\right]}\,,
\end{equation}
with $R=5.085$ fm and $5.021$ fm being the radius parameter for Ru and Zr, respectively,  and $a=2.3/(4\ln(3))=0.523$ fm
for both~\cite{Fricke:1995zz} (the numbers are also tabulated in Table~\ref{tab:WSSkinHalo}). 
The simplest approach is to take the proton and neutron WS parameters ($R,a$) to be as same as those for the charge density; the neutron skin thickness is implicitly zero.

\begin{figure*}
	\begin{centering}
		\includegraphics[scale=0.3]{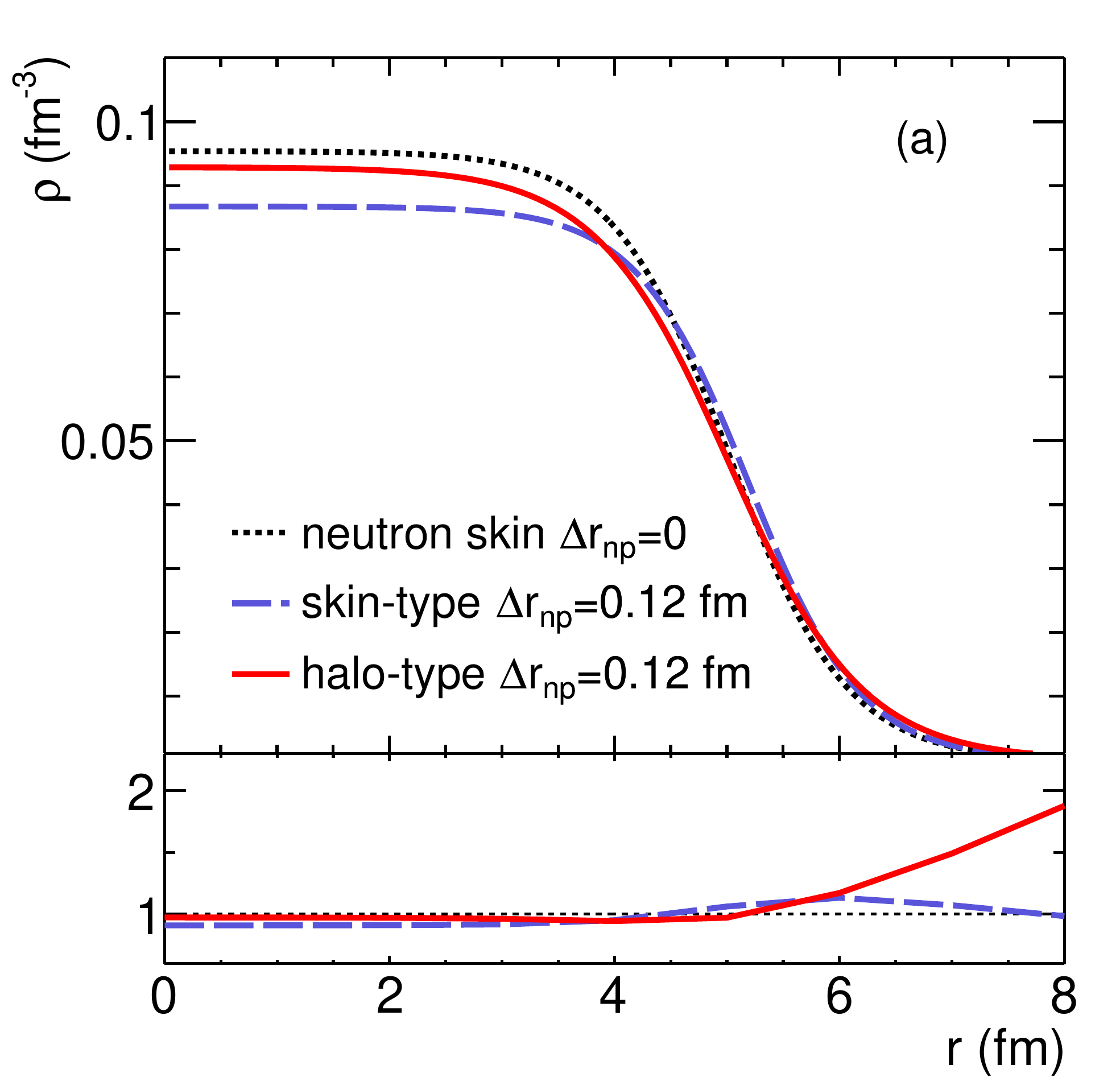}\includegraphics[scale=0.3]{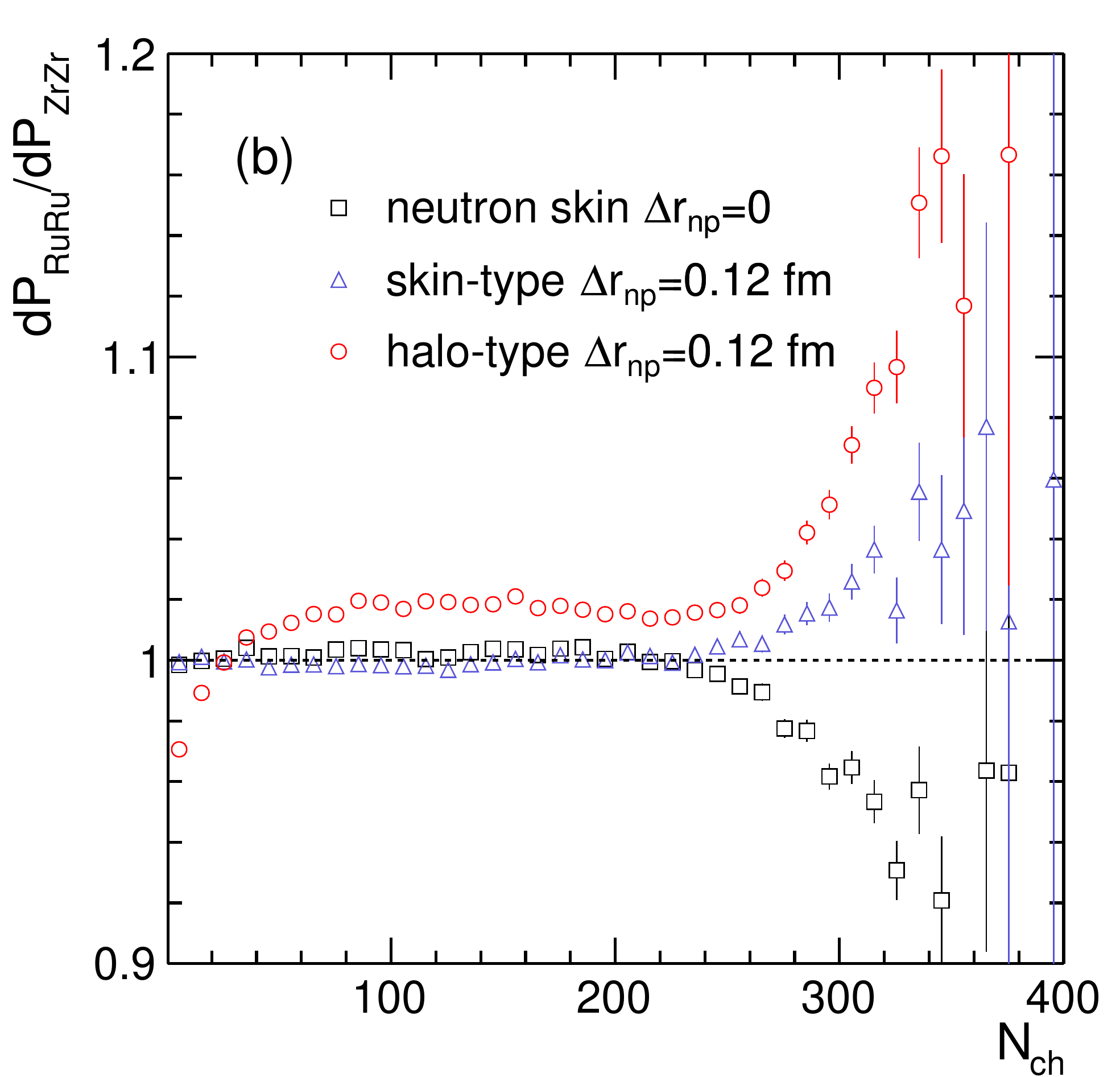}\includegraphics[scale=0.3]{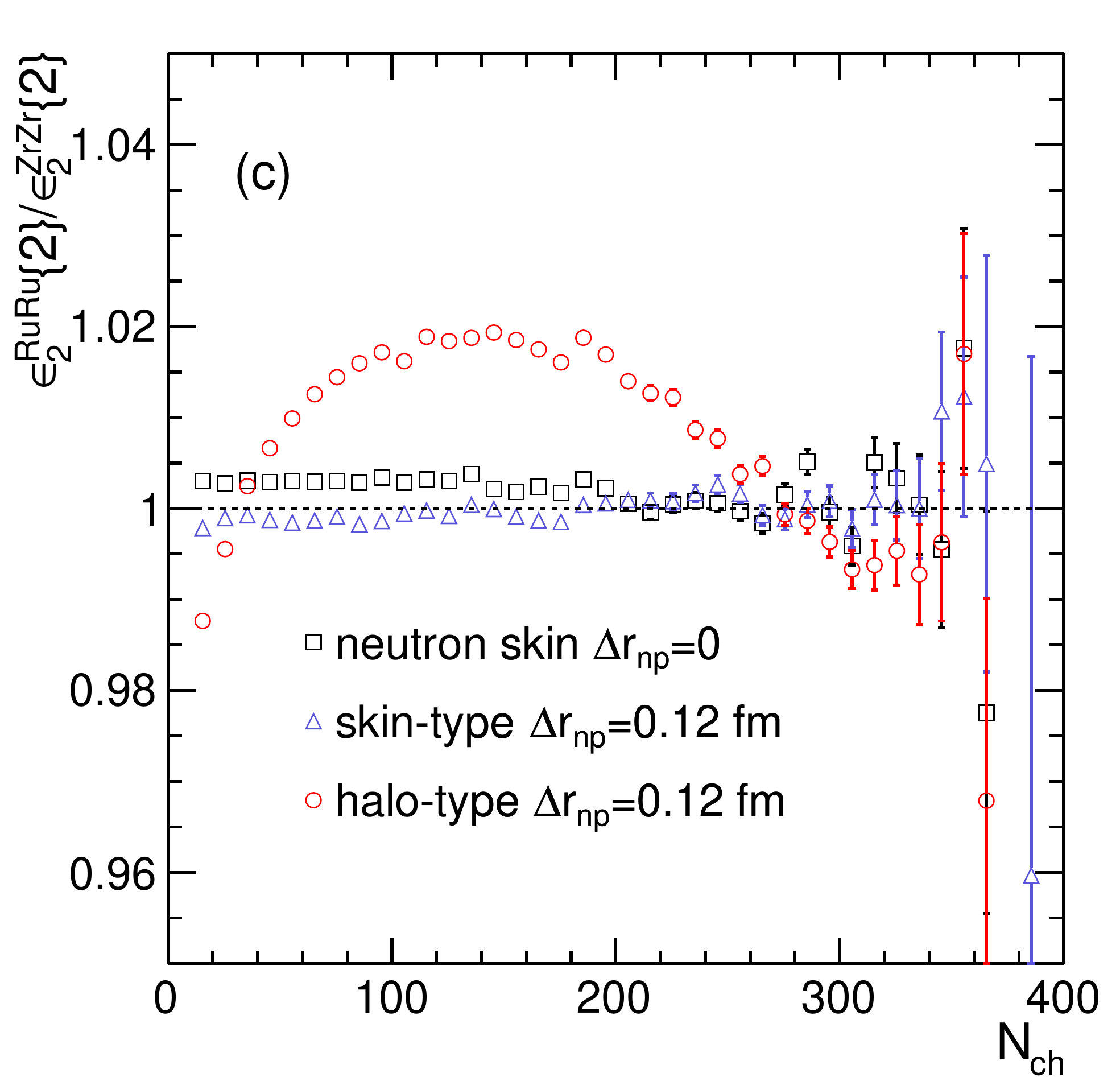}
	\par\end{centering}
	\caption{(a) The Woods-Saxon (WS) neutron density distributions of \Zr\ for
	zero neutron skin thickness, skin-type and halo-type neutron skin thickness of \rnp=0.12~fm. The lower panel shows the ratios of the densities with finite \rnp\ (with the corresponding line styles) to that without. 
	The corresponding \RuZr\ ratios of (b) charged hadron multiplicity ($\Nch$) distributions and (c) eccentricity ($\ecc$) in relativistic isobar collisions from \trento\ simulations using the three neutron densities in (a).\label{fig:WSSkinHalo}}
\end{figure*}

Figure~\ref{fig:WSSkinHalo} shows the charge density distribution (the dotted curve in panel a), the \RuZr\ ratio of the $\Nch$ distributions (the open squares in panel b) and the $\ecc$ ratio (the open squares in panel c) in Ru+Ru over Zr+Zr
collisions. It is found that the effect of the finite $R$ difference between Ru and Zr is mostly small. The large effect (bending from unity down to zero) in the $\Nch$ distribution ratio at high $\Nch$ is due to the fact that the effective
radius of Zr is smaller than that of Ru, 
which makes the overlap region denser (larger $T_A$) yielding a larger probability of high $\Nch$ events in central Zr+Zr than Ru+Ru collisions.
Similar result was found by Ref.~\cite{Deng:2016knn} where the charge WS densities were used.
The effect of the $R$ difference on the $v_2$ ratio is less than $0.5\%$. 

\subsection{Effect of neutron skin}

Intermediate to heavy nuclei have sizeable neutron skin thicknesses, $\Delta r_{\rm np}=\mean{r_{\rm n}^{2}}^{1/2}-\mean{r_{\rm p}^{2}}^{1/2}$,
where $\langle r^{2}\rangle^{1/2}$ is root-mean-square radius of either the proton or neutron distribution,
\begin{equation}\label{eq:r}
	\langle r^{n}\rangle\equiv\frac{\int r^{n}\rho(\mathbf{r})d\mathbf{r}}{\int\rho(\mathbf{r})d\mathbf{r}}\,.
\end{equation}
There are two extreme cases to constrain the WS parameters 
($\Rn,\an$) for the neutron density with finite
\rnp~\cite{Trzcinska:2001sy}: (i) skin-type density where $\Rn>\Rp$
and $\an=\Ap$; and (ii) halo-type density where $\Rn=\Rp$
and $\an>\Ap$. 
Experimental data suggest a neutron skin thickness of $\Delta r_{\rm np}=0.12\pm0.03$ fm for \Zr\ and it is of the halo-type~\cite{Trzcinska:2001sy,Jastrzebski:2004yn}; there may be additional significant theoretical uncertainty on its magnitude~\cite{Tsang:2012se}. 
We take \rnp=0.12~fm for \Zr\ and keep it zero for Ru, and investigate whether relativistic heavy ion collisions can distinguish between the two types.
If so, then it would add significant strength to our nuclear structure knowledge as  relativistic isobar collisions are distinctively different from traditional methods.
The corresponding Zr neutron density distributions are
 shown in Fig.~\ref{fig:WSSkinHalo}(a) by the dashed blue and solid red curves.

\begin{table}
	\caption{Skin- and halo-type WS parameterizations (radius parameter $R$ and diffuseness parameter $a$) of neutron densities based on the measured charge (proton) densities~\cite{Fricke:1995zz} together with neutron skin thicknesses (taken to be \rnp=0 and 0.12~fm~\cite{Trzcinska:2001sy,Jastrzebski:2004yn}, respectively) for \Ru\ and \Zr. All quoted numbers are in fm.
	\label{tab:WSSkinHalo}}
	\centering{}%
	\begin{tabular}{p{2.2cm}p{1cm}p{1cm}p{1cm}p{1cm}}
		\hline 
		& \multicolumn{2}{c}{\Ru} & \multicolumn{2}{c}{\Zr}  \\
		& $R$ & $a$ & $R$ & $a$ \\
		\hline 
		p&  5.085 & 0.523 &  5.021 & 0.523  \\
		skin-type n&  5.085 & 0.523 &  5.194 & 0.523  \\
		halo-type n&  5.085 & 0.523 &  5.021 & 0.592  \\
		\hline 
	\end{tabular}
\end{table}

The $\Nch$ and $\ecc$ ratios from \trento\ simulations using the skin-type neutron densities are shown by the blue triangles in Fig.~\ref{fig:WSSkinHalo}(b) and (c), respectively. These ratios flip about unity from those with the charge WS densities in the previous section.
This is easy to understand because the neutron skin has made Zr larger than Ru, opposite to that from the charge WS density (see Table~\ref{tab:WSSkinHalo}). 
The now smaller Ru+Ru than Zr+Zr collisions produce more particles in central collisions, making the \RuZr\ ratio larger than unity.
As previously discussed the strongest effect is on the tail of the $\Nch$ distribution.
The effects are minor on $\ecc$ and the $\Nch$ distribution in non-central collisions.
Similar results have also been found in our previous study with AMPT~\cite{Li:2018oec}.

The halo-type neutron density, on the other hand, 
has profound effects on the $\Nch$ and $\ecc$ 
in mid-central isobar collisions. This is shown in Fig.~\ref{fig:WSSkinHalo}(b) and (c) by the red circles. The increasing $\Nch$ ratio in central collisions comes from the smaller effective radius of Ru than Zr, similar to the case of skin-type neutron density. 
The ratios in mid-central collisions are distinct, with an arch shape reaching as large as $2\%$ on the top and dipping below unity in peripheral collisions.
This feature will be discussed further in the next section.

\subsection{DFT-calculated densities}

\begin{figure*}
	\begin{centering}
		\includegraphics[scale=0.3]{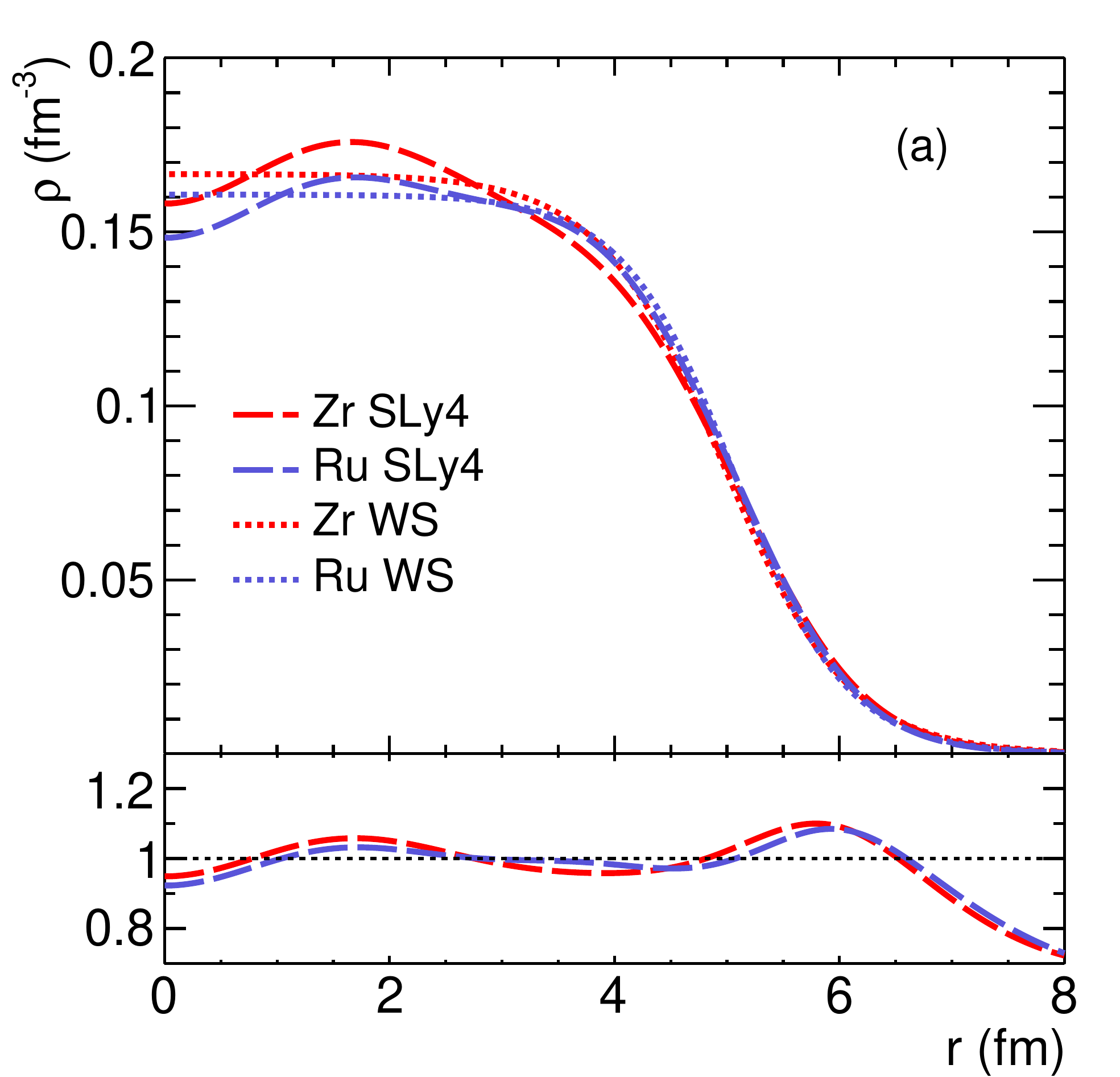}\includegraphics[scale=0.3]{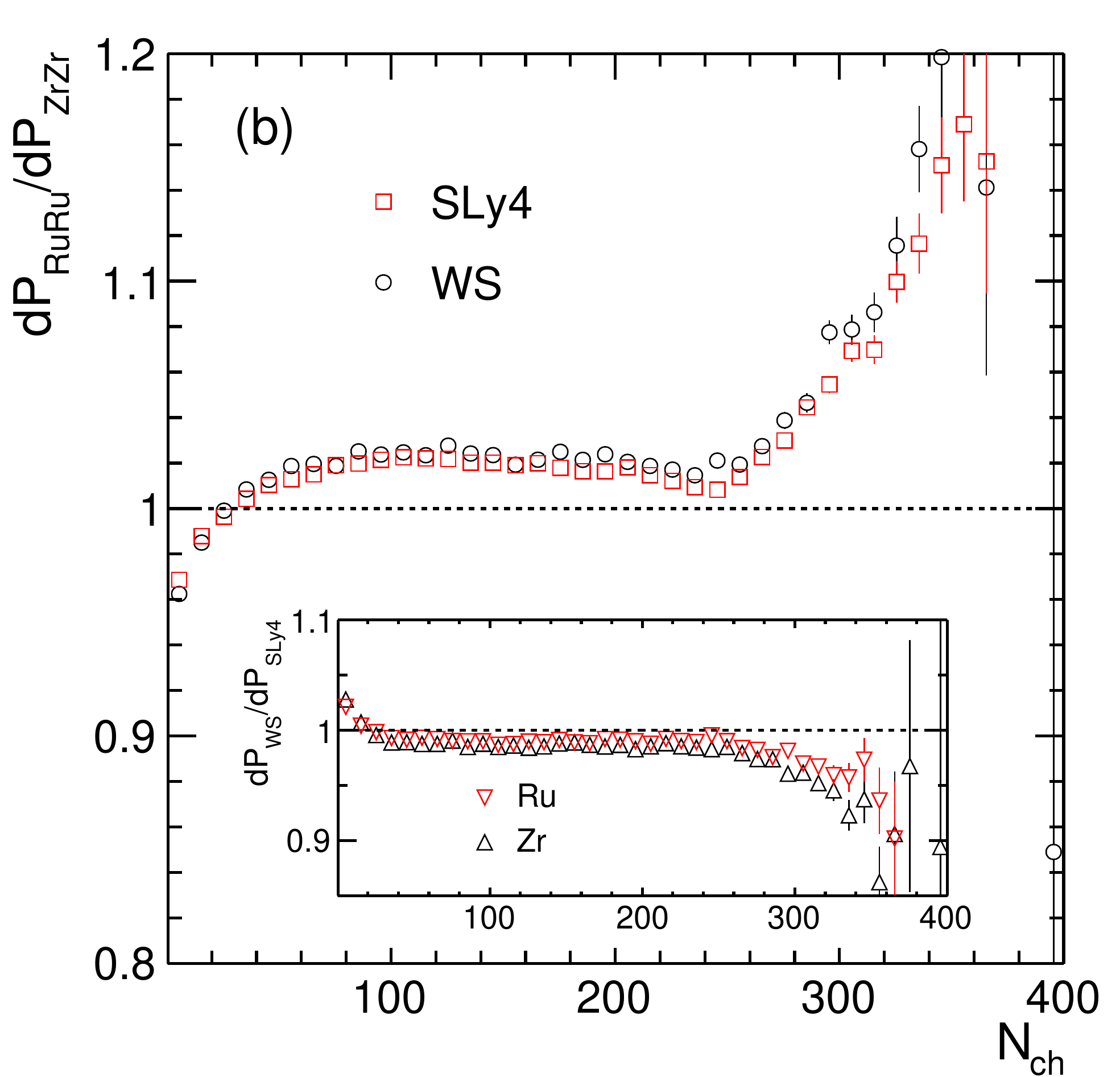}\includegraphics[scale=0.3]{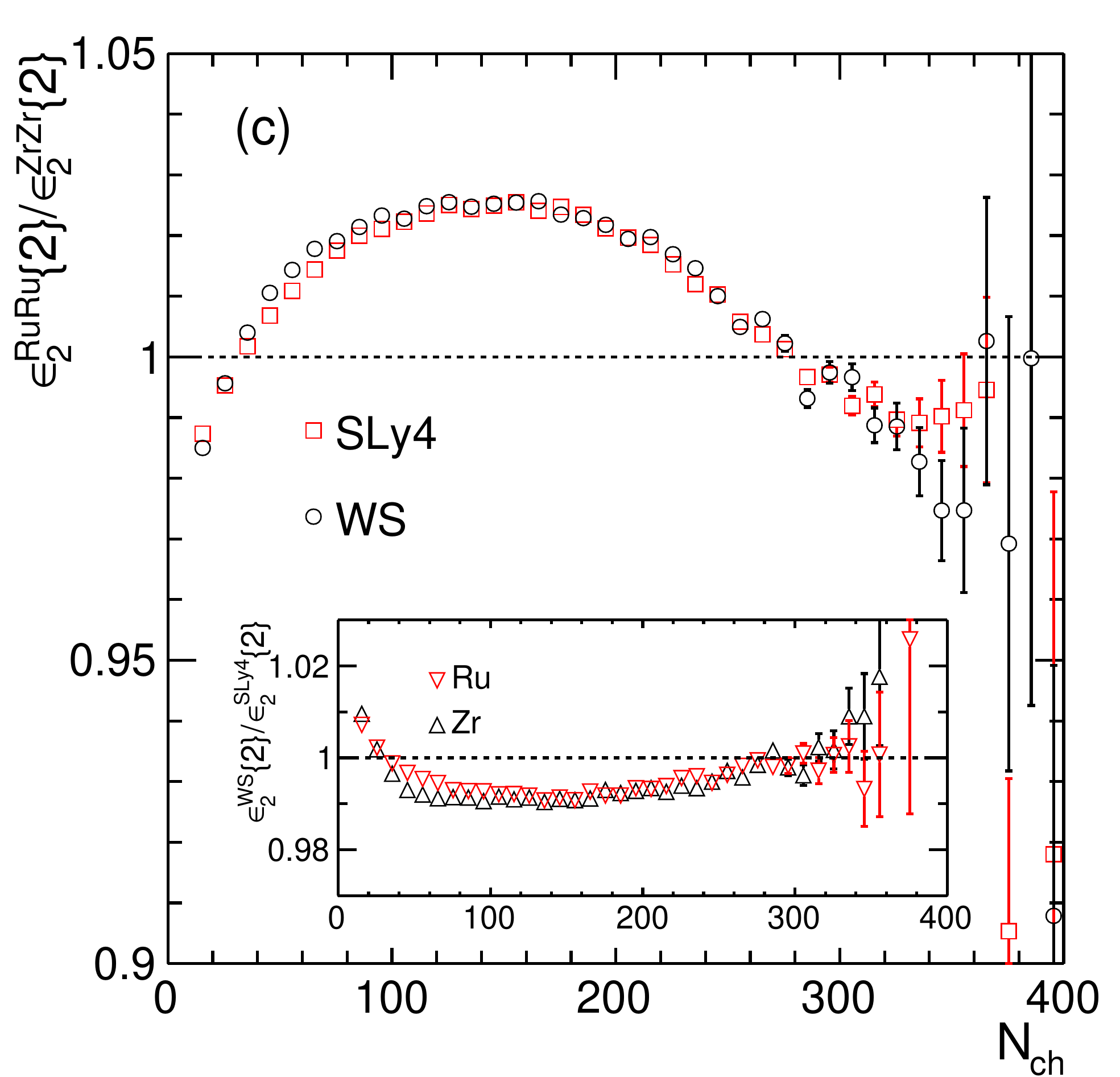}
	\par\end{centering}
	\caption{(a) The nucleon density distributions for \Ru\ and \Zr\ from energy density functional theory (DFT) calculations with SLy4 skyrme parameter set, and their corresponding Woods-Saxon (WS) parameterizations. The \RuZr\ ratios of (b) the $\Nch$ distributions  and (c) the $\ecc$ in relativistic isobar collisions with the DFT-calculated densities and the corresponding WS parameterizations. 
	The relative difference between results from the two densities in a single collision system is shown in the insets.
	\label{fig:WSDFT}}
\end{figure*}

The skin- and halo-type neutron densities are two extreme parameterizations; the truth is likely  in-between.
Nowadays, DFT is the state-of-the-art in calculating nuclear structures~\cite{Chabanat:1997qh,Chamel:2009yx,Zhang:2015vaa,Wang:2016rqh}. 
Figure~\ref{fig:WSDFT}(a) shows the DFT densities of  Zr  in dashed curves, calculated with the well known skyrme parameter set SLy4~\cite{Chabanat:1997qh} (the corresponding individual neutron and proton densities have already been shown in Ref.~\cite{Xu:2017zcn}).
Figure~\ref{fig:WSDFT}(b) and (c) show the \RuZr\ ratios of the $\Nch$ distributions and $\ecc$, respectively, simulated by the \trento\ model taking the DFT densities as input.
It is immediately clear that the ratios are similar to those using halo-type neutron densities in Fig.~\ref{fig:WSSkinHalo}.
Similar features have also been observed with Glauber as well as AMPT calculations taking DFT densities as input~\cite{Xu:2017zcn,Li:2018oec}.
This suggests, model-independently, that the DFT densities may have halo-type neutron skin.
In Ref.~\cite{Li:2018oec} we have tested a skin-type WS density keeping $a$ fixed and adjusting $R$ to the DFT-calculated density via $\mean{r^2}$ of Eq.~\ref{eq:r}. The DFT-adjusted skin-type WS density yielded results similar to the skin-type neutron density results in Fig.~\ref{fig:WSSkinHalo}(b) and (c). This indicates that the DFT density cannot be substituted by a skin-type density even if the effective radius is forced to be the same.

The DFT density contains more detailed information than the WS parameterization. 
Many of those details may not be important for a given observable of interest.
The question arises whether a proper WS approximation of the DFT density would be sufficient for our $\Nch$ and $\ecc$ studies.
Clearly, skin- and halo-type neutron densities do make significant differences as shown in Fig.~\ref{fig:WSSkinHalo}, so not any parameterization would work.
We can determine both WS parameters $R$ and $a$  by matching the $\mean{r}$ and $\mean{r^2}$ from the DFT proton and neutron densities~\cite{Dobaczewski:1994zza}. 
The obtained parameters are tabulated in Table~\ref{tab:WSDFT}.
The parameters do indicate that the DFT density is 
closer to the halo-type WS density ($\Rn\approx\Rp$ but $\an>\Ap$).
This is in line with the findings by low-energy nuclear experiments~\cite{Trzcinska:2001sy,Jastrzebski:2004yn}. 

\begin{table}
	\caption{The WS parameterizations (radius parameter $R$ and diffuseness parameter $a$) of proton and neutron (and nucleon) density distributions for \Ru\ and \Zr, matching to the corresponding $\mean{r}$ and $\mean{r^2}$ from the DFT-calculated spherical densities with SLy4 skyrme parameter set~\cite{Chabanat:1997qh,Xu:2017zcn}. The WS parameterization of nucleon density assuming a quadrupole deformity parameter $\beta_{2}=0.16$ and matching to the spherical DFT density is also listed. All quoted numbers are in fm. \label{tab:WSDFT}}
	\centering{}%
    \begin{tabular}%{llcccc}
    {p{1.8cm}p{1cm}p{1cm}p{1cm}p{1cm}p{1cm}}
    \hline
    & & \multicolumn{2}{c}{\Ru} & \multicolumn{2}{c}{\Zr}  \\
	& & $R$ & $a$ & $R$ & $a$ \\
    \hline
	\multirow{3}{*}{$\beta_2=0$} 
	& p   &  5.060 & 0.493 & 4.915 & 0.521   \\
	& n   &  5.075 & 0.505 & 5.015 & 0.574   \\
	& p+n &  5.067 & 0.500 & 4.965 & 0.556   \\
	\hline
	\multirow{3}{*}{$\beta_2=0.16$} 
	& p  &   5.053& 0.480 & 4.912 & 0.508   \\
	& n  &   5.073& 0.490 & 5.007 & 0.564   \\
	& p+n &  5.065 & 0.485 & 4.961 & 0.544   \\
	\hline
	\end{tabular}
\end{table}

We note that, because of the isospin symmetry of the strong interaction, most observable in relativistic heavy ion collisions, including the $\Nch$ and $v_2$ we study here, are sensitive only to the total nucleon density, the sum of the proton and neutron ones. Since what is experimentally measured is typically the charge (proton) density, how the neutron density is implemented is important. In our \trento\ model simulations, we 
use the resultant total nucleon densities; their WS parameters are also tabulated in Table~\ref{tab:WSDFT}.

Figure~\ref{fig:WSDFT}  shows the \RuZr\ ratios of the $\Nch$ distributions and $\epsilon_2$   calculated by the \trento\ model with  WS densities parameterized from the DFT-calculated ones, and compares them to those calculated directly from the DFT densities.
It is found that the DFT densities and the WS parameterized ones give essentially the same \RuZr\ ratios.
The two densities do yield slightly different results for each individual Ru+Ru or Zr+Zr system, up to a couple of percent (see the insets), but these differences cancel in the \RuZr\ ratios.
This indicates that the properly parameterized WS densities are indeed sufficient to study the $\Nch$ and $\epsilon_2$.
Note that the ratios in Fig.~\ref{fig:WSDFT} are somewhat larger than the ones from the halo-type densities in Fig.~\ref{fig:WSSkinHalo}. 
This is because the neutron skin thickness of Zr from the DFT calculation is \rnp=0.16~fm, larger than the 0.12~fm used in Fig.~\ref{fig:WSSkinHalo}.
It implies that those ratios in non-central collisions can further be used to quantitatively measure the neutron skin thickness. 
The halo-type neutron densities from DFT indicate that  Zr  has a larger diffuseness parameter $a$ and rms radii than  Ru. 
These yield distinctive shapes of \RuZr\ ratios of the $\Nch$ distributions and $\epsilon_2$.

For this study we have generated 50 million events for each collision species in each density case. Total over 2 billion events have been collected for each isobar system by the STAR experiment at RHIC~\cite{Adam:2019fbq}, so the statistical uncertainties would be negligible. 
Since most of the systematic uncertainties cancel in those ratios between the two isobar systems, our results  can be readily compared against experimental data to yield valuable information on the neutron skin type and thickness. 
If experimental data, on the other hand, do not favor our DFT and halo-type neutron density predictions, but rather the skin-type neutron density ones, 
then an overhaul of our current knowledge of nuclear structure would be called for.

\section{Effect of nuclear deformation}\label{sec:deformity}

So far, we have focused on the discussion of spherical nuclei. The elliptic
flow in most central collisions are extra sensitive to  nuclear deformation,
as  deformed nuclei colliding at zero impact parameter can have large eccentricity depending on the collision orientation~\cite{Filip:2009zz,Voloshin:2010ut}. 
To take nuclear deformation into account, the following extended WS formula~\cite{Hagino:2006fj}, 
\begin{equation}
	\rho(r,\theta)=\frac{\rho_{0}}{1+\exp\left\{ \left[r-R(1+\beta_{2}Y_{2}^{0}(\theta))\right]/a\right\} }\,,
\end{equation}
is used, where $Y_{2}^{0}(\theta)=\frac{1}{4}\sqrt{\frac{5}{\pi}}(3\cos^{2}\theta-1)$
and $\beta_{2}$ is the quadrupole deformity parameter. DFT calculations of deformed nuclei are more challenging and often result in large uncertainties on $\beta_2$. 
Experimental measurements of $\beta_2$ also suffer from large uncertainties.
\Ru\ ($\bRu = 0.158$) could be more deformed than \Zr\ ($\bZr = 0.08$), or could be less deformed ($\bRu=0.053$, $\bZr = 0.217$)~\cite{Deng:2016knn}.

\begin{figure*}
	\begin{centering}
		\includegraphics[scale=0.38]{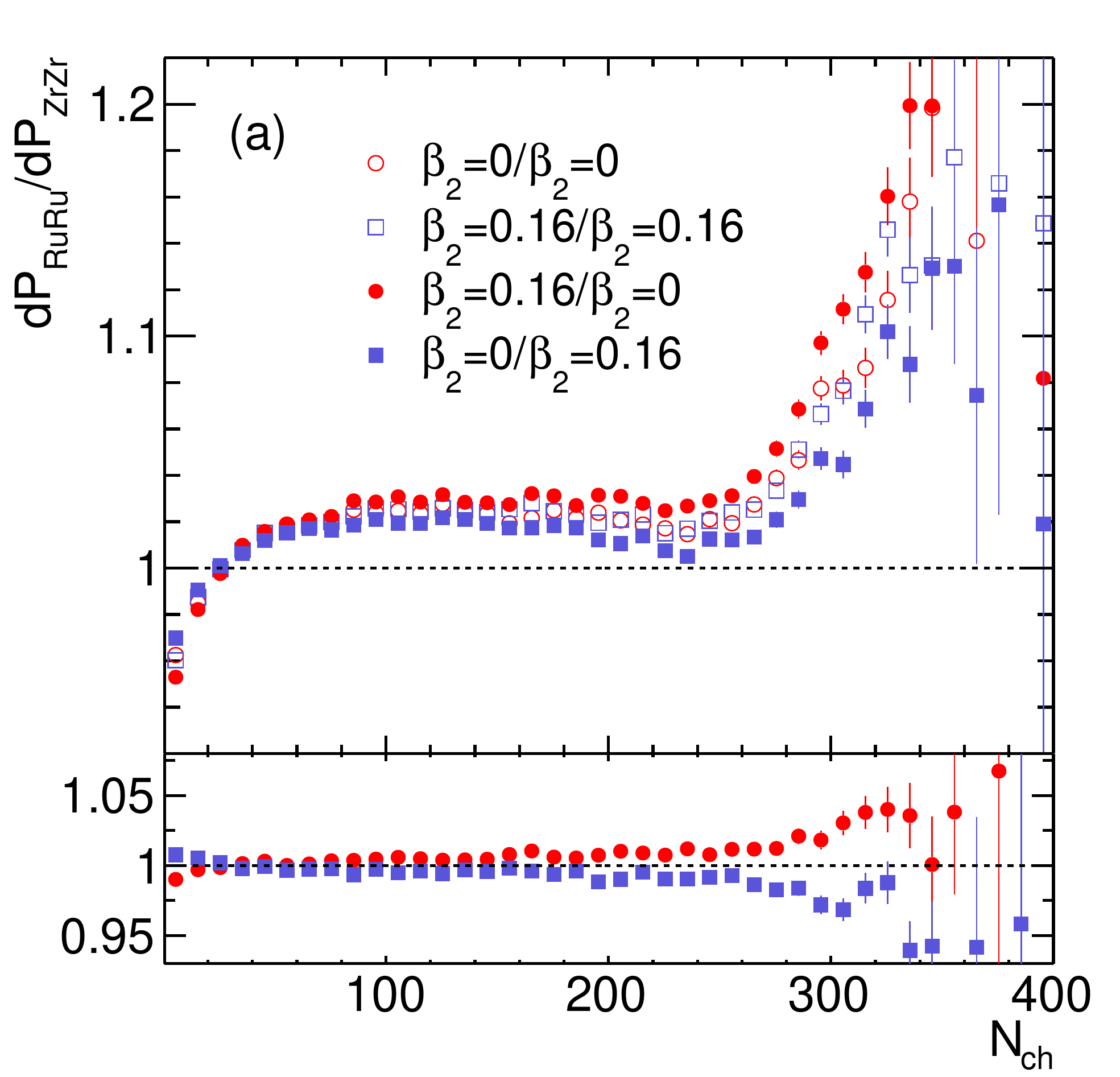}\includegraphics[scale=0.38]{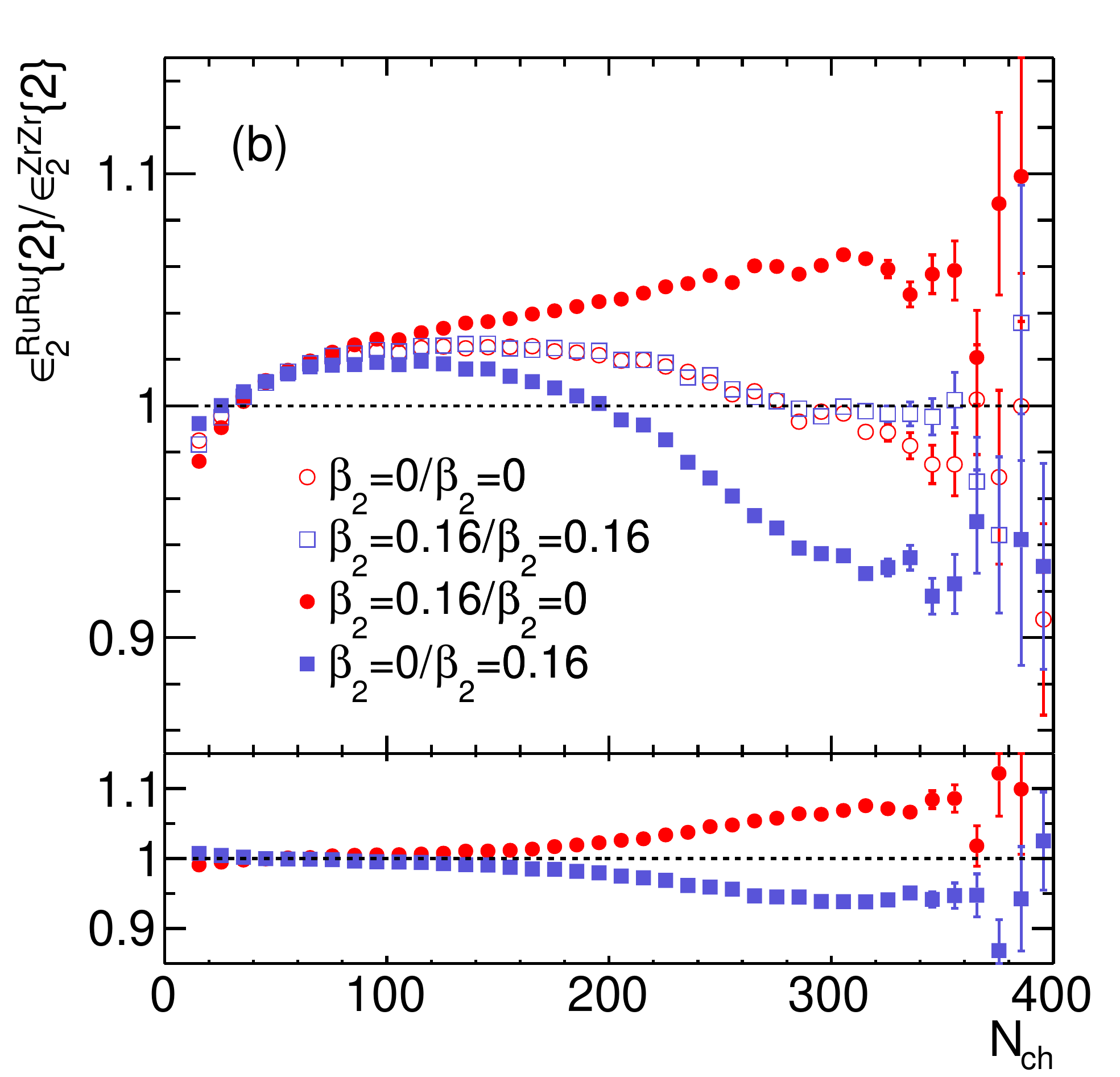}
	\par\end{centering}
	\caption{The \RuZr\ ratios of (a) the $\Nch$ distributions and (b) the $\ecc$ in relativistic isobar collisions  with various combinations of  \Ru\ and \Zr\ nuclear deformities. The lower panels show the double ratios of the deformed/spherical (and spherical/deformed) to the spherical/spherical results. \label{fig:Deformed}}
\end{figure*}

In the previous section, we have demonstrated that DFT densities can be adequately approximated by halo-type neutron densities for comparative studies of $\Nch$ and $\ecc$ in isobar collisions. 
We assume this preserves for deformed Ru and Zr; this assumption is important as it allows us to bypass the  challenging DFT calculations to still investigate efficiently the effects of nuclear deformity. 
With a given $\beta_2$, we determine $R$ and $a$ of the proton and neutron densities, respectively, by matching the $\mean{r}$ and $\mean{r^2}$ (Eq.~(\ref{eq:r})) of the corresponding spherical DFT densities.
We set $\beta_2=0.16$ for this initial study; the corresponding WS parameters are listed in Table~\ref{tab:WSDFT}. 
The finite $\beta_{2}$ reduces the $a$ parameter but have negligible 
effect on $R$.

Figure~\ref{fig:Deformed} shows the \RuZr\ ratios of the $\Nch$ distributions and $\ecc$ with various combinations of Ru and Zr deformities. 
The $\Nch$ distribution ratio changes insignificantly due to the finite
$\beta_{2}$; the change comes primarily from that in $a$, e.g.~a finite $\beta_{2}$ for Zr reduces the $a$ value yielding a larger $\Nch$ in Zr+Zr collisions compared to the  spherical case and a smaller ratio of the $\Nch$ distributions.
The $\ecc$ ratios, with the same $\beta_{2}$ value for both Ru and Zr, are almost identical for the spherical ($\beta_{2}=0$) and deformed ($\beta_{2}=0.16$) cases, except perhaps in very central collisions. 
If the two nuclei differ in $\beta_{2}$, then the $\ecc$ ratio is significantly impacted in central (and semi-central) collisions, as expected~\cite{Giacalone:2021uhj}. 
However, the effect of deformation on the $\ecc$ ratio in peripheral and semi-peripheral collisions is insignificant. 
This feature indicates that the shape of the $\ecc$ ratio in peripheral and semi-peripheral collisions can  still be used to determine the neutron skin type and thickness as we demonstrated in the previous section with spherical nuclei.
As aforementioned, 40 times more statistics have been collected by STAR than used in this simulation~\cite{Adam:2019fbq}. Even though experimental triggers suffer inefficiencies in peripheral collisions, it is significant only in very peripheral collisions ($\Nch\lesssim 10$), and even there the inefficiency is larger than 10\%. 
In other words, despite the large bias to the $v_2$ ratio in (semi-)central
collisions, our proposal to discriminate the neutron skin by \RuZr\ ratios of $\Nch$ distributions and $v_2$ is still feasible even though the Ru and Zr deformities are largely uncertain.
Furthermore, the $v_2$ ratio in central collisions, together with $\Nch$ distribution ratio, can be exploited to probe the $\beta_{2}$ parameters of the isobar nuclei; we postpone such a study to further work. 

\section{Summary and outlook}\label{sec:summary}
By using the \trento\ model, we have investigated the effects of neutron skin  on the ratios of the charged hadron multiplicity ($\Nch$) distributions	and eccentricity ($\ecc$)  in relativistic \RuRu\ over \ZrZr\  collisions at $\snn=200$ GeV.
It is found that these \RuZr\ ratios are exquisitely sensitive to the neutron skin type (skin vs.~halo) and thickness. Taking the Woods-Saxon (WS) neutron densities to be as same as the proton densities for both nuclei (i.e.~without any neutron skin), the \RuZr\ ratio of the $\Nch$ distributions bends down below unity at large $\Nch$ because of the smaller effective Zr radius than the Ru's in these charge distributions. Including neutron skin, which is thicker in Zr than in Ru, the ratio bends up above unity at large $\Nch$ because the effective nucleon radius of Zr is now larger than the Ru's. The shapes of the \RuZr\ ratios of the $\Nch$ distributions and $\ecc$ in mid-central collisions can further distinguish between skin-type and halo-type neutron densities, both having the same skin thickness (see Fig.~\ref{fig:WSSkinHalo}). 

The \RuZr\ ratios obtained by using nuclear densities from the state-of-the-art calculations by energy density functional theory (DFT) show similar features as those using nuclear WS density distributions with a halo-type neutron skin (see Fig.~\ref{fig:WSDFT} and Ref.~\cite{Xu:2017zcn,Li:2018oec}). This is because the DFT-calculated densities are more like the halo-type than the skin-type (see Table~\ref{tab:WSDFT}).
The soon-to-be-available isobar data should be able to determine whether the neutron skin is skin-type or halo-type, how large the neutron skin thickness (\rnp) is, and whether or not the DFT densities are a truthful reflection of nuclear structures.

After demonstrating that the halo-type WS densities can substitute the DFT-calculated ones to faithfully describe the \RuZr\ ratios, we include a nuclear quadrupole deformity into the WS distribution of Ru and/or Zr to study the effects.
We show, as a proof of principle, that the ratios of the $\Nch$ distributions and $v_2$  in isobar collisions can be used to probe the nuclear deformation (see Fig.~\ref{fig:Deformed}).
Corroborating with other means of probing nuclear deformity, e.g.~$v_2$-$\mean{p_T}$ correlations~\cite{Giacalone:2019pca}, should yield valuable information on nuclear structure in the future. 

\section*{Acknowledgments}
This work is supported in part by the Zhejiang Provincial Natural Science Foundation of China (Grant No.~LY21A050001), the National Natural Science Foundation of 
China (Grant Nos.~11905059, 11947410, 12035006, 12047568, 12075085), the Ministry of Science and Technology of China (Grant No.~2020YFE0202001), and the U.S.~Department of Energy (Grant No.~DE-SC0012910).

\bibliographystyle{elsarticle-num}
\bibliography{ref}
\end{document}